\documentclass[9pt,twocolumn,twoside]{osajnl}

\journal{ol} 

\setboolean{shortarticle}{true}

\title{Multimode Mamyshev Oscillator}

\author[1,*]{Henry Haig}
\author[1]{Pavel Sidorenko}
\author[2]{Anirban Dhar}
\author[2,3]{Nilotpal Choudhury}
\author[2]{Ranjan Sen}
\author[4]{Demetrios Christodoulides}
\author[1]{Frank Wise}

\affil[1]{School of Applied and Engineering Physics, Cornell University, Ithaca, New York 14853, USA}
\affil[2]{Fiber Optics and Photonics Division, CSIR-Central Glass and Ceramic Research Institute, Kolkata-700032, India}
\affil[3]{Academy of Scientific and Innovative Research (AcSIR), Gaziabad 21002, India}
\affil[4]{CREOL/College of Optics and Photonics, University of Central Florida, Orlando, Florida 32816, USA}
\affil[*]{Corresponding author: tsh67\@cornell.edu}

\begin{abstract}
We present a spatiotemporally mode-locked Mamyshev oscillator. A wide variety of multimode mode-locked states, with varying degrees of spatiotemporal coupling, are observed.  We find that some control of the modal content of the output beam is possible through the cavity design. Comparison of simulations to experiments indicates that spatiotemporal mode-locking is enabled by nonlinear intermodal interactions and spatial filtering, along with the Mamyshev mechanism. This work represents a first exploration of spatiotemporal mode-locking in an oscillator with the Mamyshev saturable absorber.
\end{abstract}

\setboolean{displaycopyright}{true}

\begin{document}

\maketitle

\section{Introduction}
Mode-locked fiber lasers attract considerable interest in science, industry, and healthcare due to their compactness, efficiency, and inherent alignment-free design. However, fiber lasers lag behind their solid-state counterparts in pulse energy owing mainly to strong nonlinear processes in the waveguide medium. While nonlinearity enables mode-locking, it eventually destabilizes pulse evolutions and constrains performance. Consequently, researchers explore methods to benefit from nonlinearity as well as ways to reduce nonlinear effects.

One recent innovation in exploiting nonlinearity is use of the Mamyshev mechanism \cite{mamyshev} in fiber oscillators. Mamyshev oscillators contain two offset spectral filters \cite{Rochette, Piche, Regelskis:15} that extinguish continuous-wave (CW) lasing, but allow pulses with sufficient peak power to traverse the cavity through nonlinear spectral broadening between filters. This enacts strong effective saturable absorbtion \cite{Pitois:08} which can stabilize highly nonlinear evolutions. Mamyshev oscillators achieve pulse energies (> 200 nJ in single-mode fiber \cite{Sidorenko:18}, > 1 µJ in photonic crystal fiber \cite{Liu:19}) and durations (< 40  fs) \cite{Sidorenko:18} far beyond those of fiber lasers with other saturable absorbers.

Increasing fiber core size is another technique to scale energy, by reducing nonlinearity. This method is limited in principle for single-mode (SM) lasers by the higher-order modes that exist in large-core fiber. However, the demonstration of spatiotemporal mode-locking (STML) in multimode (MM) fiber \cite{Wright94} showed that it is possible to incorporate higher-order modes in mode-locking. The limits of STML are unknown, and the broad range of nonlinear multimode propagation phenomena \cite{krupa} may provide means to generate light with unprecedented degrees of spatiotemporal control. Some STML lasers build on well-known SM phenomena, such as dissipative soliton \cite{Wright94} and self-similar evolutions \cite{Tegin:19}. Others exploit spatiotemporal degrees of freedom in entirely new ways, such as beam self-cleaning \cite{teginbc}. 

Much remains to be explored in MM oscillators. Previous work on STML has highlighted the importance of linear filtering and nonlinear intermodal interactions, but the effect of saturable absorption on MM pulse evolution is not fully understood. To date, STML lasers have used saturable absorbers based on material absorption \cite{materialsa} or nonlinear polarization evolution  \cite{STIND, Tegin:19, Wright94, teginbc}. The Mamyshev saturable absorber has not been applied to STML. Most generally, one would expect that a saturable absorber with deep modulation would be attractive for STML. However, it is unknown how the Mamyshev mechanism will affect MM evolutions in general, and whether it can provide similar or alternative benefits to MM oscillators as it has to SM oscillators.

Here we present a first demonstration of spatiotemporal mode-locking in a Mamyshev oscillator. The laser exhibits a wide variety of STML states, ranging from compressible single pulses to spatiotemporally-complex pulse bursts. In some states, the output profile can be partially controlled by the cavity alignment. Numerical simulations that account for the main features of the experimental observations will also be presented.

\begin{figure}[h!]
\centering
\includegraphics[width=\linewidth]{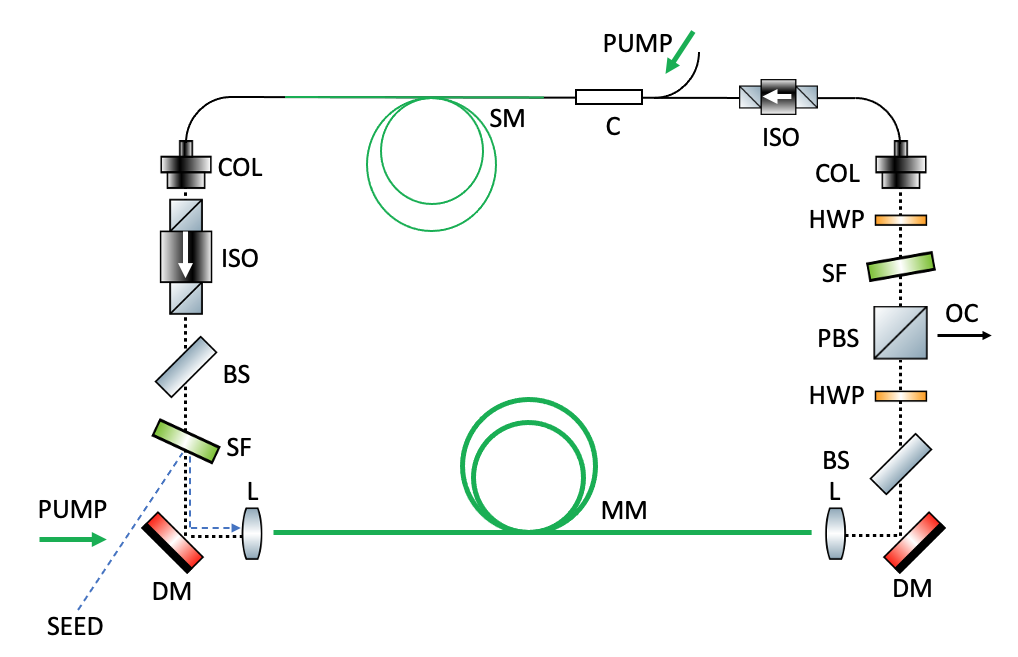}
\caption{Cavity Schematic of the Multimode Mamyshev Oscillator. DM, dichroic mirror; MM, multimode fiber; L, lens; BS, beam sampler; HWP, half-wave plate; PBS, polarizing beam splitter; OC, output coupling; SF, spectral filter; COL, collimating lens; ISO, isolator; C, combiner; SM, single mode fiber. Green indicates Yb doped gain fiber, and black indicates passive fiber.}
\label{fig:schematic}
\end{figure}

Fig. \ref{fig:schematic} is a schematic of the laser cavity. The ring cavity comprises one SM arm and one MM arm separated by offset spectral filters (4 nm full-width-at-half-maximum (FWHM) bandwidth and 5 - 10 nm center wavelength separation). The filters, along with nonlinear spectral broadening in either arm, create an effective saturable absorber. The SM arm includes 2 meters of passive fiber and 1.5 meters of active fiber. Both segments are polarization-maintaining (PM) to suppress polarization dynamics. 8.5 µm core diameter fiber is used to deliver sufficient average power to the MM arm to saturate the gain in that segment. The multimode arm consists of 1.5 - 2 m of 30 µm core diameter Yb-doped graded-index fiber. GRIN MM fiber is chosen for its small intermodal group velocity differences. Both gain fibers are pumped by wavelength-stabilized MM pump diodes at 976 nm. Waveplates and a beam splitter provide MM output coupling and align the polarization with the PM fiber's axis. These design choices were motivated by the goal of limiting the complexity of the evolution, as observed in initial simulations of the cavity.

The GRIN MM fiber was fabricated through the vapor phase chelate delivery technique \cite{Choudhury} by optimization of different fabrication parameters such as vapor-phase composition, sublimator temperature, deposition temperature, and number of deposition passes. The vapor phase composition was specifically adjusted in different core layers to achieve a graded profile along the radial direction which finally yielded a fiber with numerical aperture of 0.14 and maximum Yb-ion-concentration of 4000 ppm. The fiber supports 20 transverse modes of a single linear polarization at 1030 nm according to our mode calculations \cite{waveguide}.

For the experiments described here, the oscillator is started by seeding with pulses from an ANDi fiber oscillator \cite{Chong:06} with 20 nm bandwidth centered at 1040 nm and 1.3 ps (chirped) duration. The seed pulses are coupled to the cavity by reflection from the spectral filter preceeding the MM arm. While aligning the laser, we measure the spatially-integrated MM beam spectrum with a diffuser and highly-MM fiber coupled spectrometer. The temporal behavior is monitored with a photodiode and oscilloscope, and the near-field MM beam profile is measured with a camera. When mode-locking occurs, the mode-locked spectrum and pulse train dominate over the seed spectrum and pulse train. The seed beam is then blocked for self-sustaining pulsed operation. 

\begin{figure}[htbp]
\centering
\includegraphics[width=\linewidth]{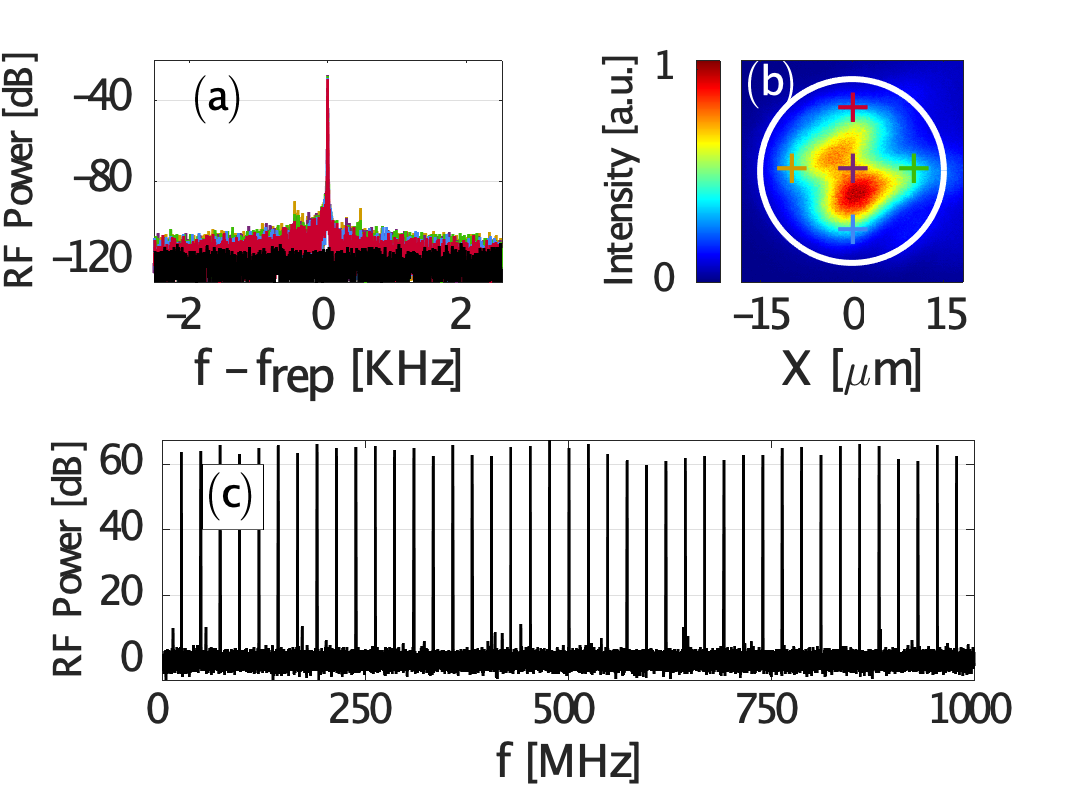}
\caption{Typical RF spectra of an STML state. (a) shows overlapping RF spectra signal spikes at the cavity fundamental repetition rate (23 MHz) within a 5 KHz window for different parts of the beam. Colored data correspond to the crosshairs in (b). Black curve is the background noise floor. (b) shows the near-field beam profile. The white circle is the core-cladding boundary. The intensity colormap and scale here are used for all other images presented in this letter. (c) shows the RF spectrum of the center section of the beam with harmonics up to the response limit of the photodetector (1 GHz).}
\label{fig:RF}
\end{figure}

To characterize the laser, we use a pinhole to isolate spatial sections of the beam (< 10\% of the beam) for measurements. The laser produces a wide variety of pulses with various spectral bandwidths (7 – 30 nm), spatial profiles, energies (7 – 20 nJ), and spatiotemporal complexity. In every mode-locked state we observe, the RF spectrum is close to uniform across the beam, as shown in Fig. \ref{fig:RF}. Contrast between the fundamental frequency and secondary modulations over 70 dB indicates stable mode locking at the fundamental cavity repetition rate. We also observe some states for which the RF spectrum and photodiode trace show multi-pulsing, albeit always uniformly across the beam. We use autocorrelations based on second-harmonic generation (SHG) to determine the pulse duration. SHG autocorrelations generally cannot accurately measure spatiotemporally-coupled beams \cite{Wright2020} because of the limited angular acceptance bandwidth. However, we find that sufficiently small sections of the beam give autocorrelations that are insensitive to the phase-matching condition, and so can be measured with this technique. The output pulses are chirped, with spatially-integrated duration around 1 ps. All mode-locked states that we observe so far exhibit spatio-spectral complexity, meaning the optical spectrum varies across the beam. Below, we present two representative STML states, with minimal and maximal spatiotemporal coupling, respectively.



\begin{figure}[h!]
\centering
\includegraphics[width=\linewidth]{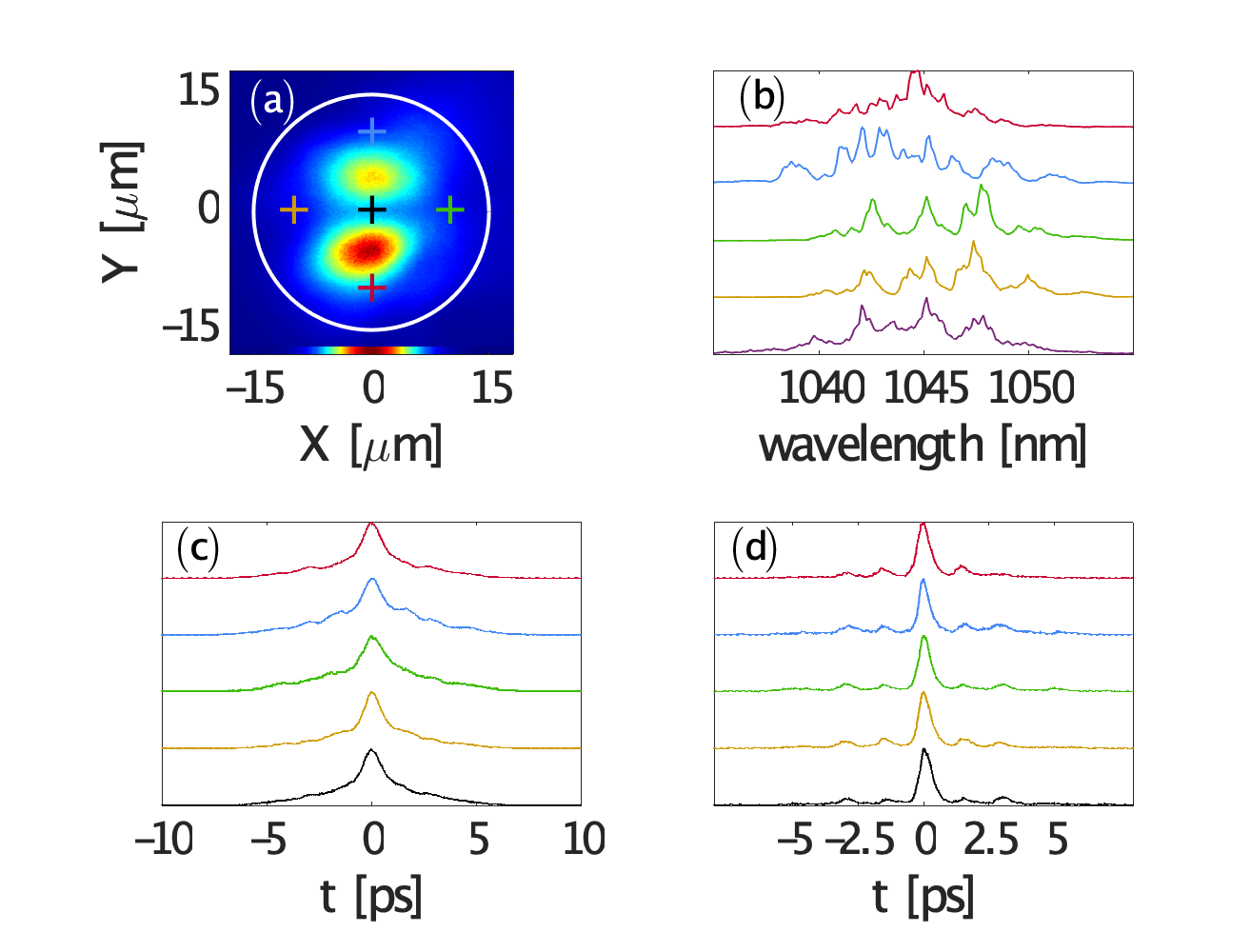}
\caption{Measurements of 15-nJ pulses with weak spatiotemporal coupling. (a) Near-field beam profile with core boundary in white and fundamental mode cross section inset at bottom. Measurements in (b), (c), and (d) are taken by isolating sections of the beam indicated by corresponding colored crosshairs in (a). (b) normalized spectra, with the spatially-integrated spectrum in purple. (c) Autocorrelations of pulses directly from laser. (d) Autocorrelation of compressed pulses. The optimal dispersion for compression is the same for each section.}
\label{fig:meascompressible}
\end{figure}

Measurements of a single-pulse mode-locked state with minimal variation of the temporal profile across the beam are summarized in Fig. \ref{fig:meascompressible}. Despite spatio-spectral variation, the autocorrelation is nearly identical for any section of the beam. These pulses compress reasonably well by a standard grating compressor to a FWHM duration of 300 fs, with a transform-limited FWHM duration of 200 fs. We estimate that under 20\% of the total pulse energy is contained in pedestal structure.  As a secondary point, the pulses from the SM arm (20 nJ with 60 nm FWHM bandwidth) are similar to those from SM Mamyshev oscillators \cite{Sidorenko:18} \cite{Sidorenko:19}.

\begin{figure}[h!]
\centering
\includegraphics[width=\linewidth]{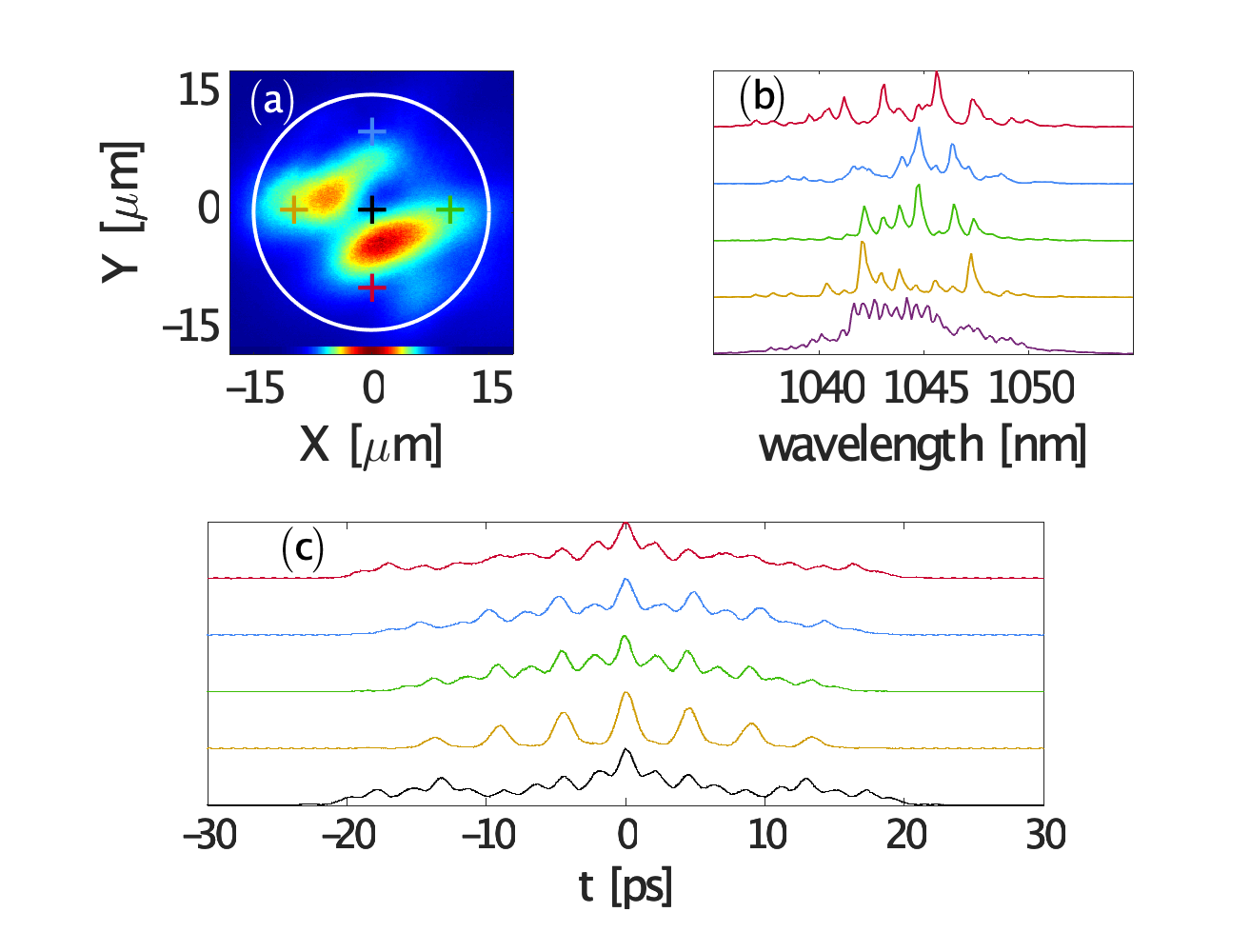}
\caption{Measurements of a highly-spatiotemporally-coupled pulse burst of 20 nJ total energy. (a), (b), and (c) correspond to the same measurements as in Fig. \ref{fig:meascompressible}.}
\label{fig:stc}
\end{figure}

In other cases, such as the state shown in Fig. \ref{fig:stc}, the autocorrelation varies considerably with the chosen portion of the beam, which is evidence of strong spatiotemporal coupling. Typically, these states exhibit multiple pulses with picosecond separation, and the effect of a grating compressor depends strongly on the chosen section of the beam. These states may be undesirable from an applications standpoint, since they would require complex spatiotemporal phase compensation to compress.


Attempts to increase the pulse energy by increasing pump power usually result in formation of multiple pulses with picosecond separation. Increasing filter separation is an established technique to prevent multipulsing or increase the single-pulse energy of Mamyshev oscillators \cite{Chen:21} \cite{Zeludevicius:18} by increasing the saturation power of the saturable absorber. In this laser, however, we find that increasing the filter separation past 10 nm disrupts mode-locking before it prevents multi-pulsing. 

\begin{figure}[htbp]
\centering
\includegraphics[width=\linewidth]{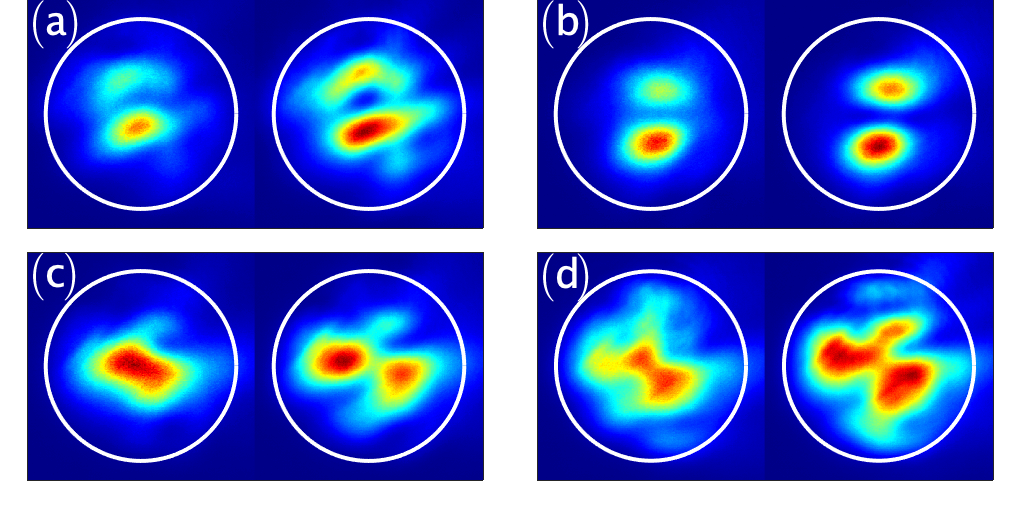}
\caption{Each pair of images shows the multimode near field beam profile before (left) and after (right) mode-locking.}
\label{fig:baml}
\end{figure}

The beam profile prior to mode-locking is determined by the coupling of filtered ASE from the SM arm to the MM arm. We observe qualitatively that this coupling usually determines the spatial profile of the mode-locked state. This trend is shown in Fig. \ref{fig:baml}(a) and (b), where the spatial profile before seeding and mode-locking strongly resembles that after seeding and mode-locking. Less often, we observe states for which the profiles are distinct (Fig. \ref{fig:baml}(c) and (d)). This phenomenon is not fully understood, but we speculate that the modal energy distribution in the MM arm is primarily determined by the SM to MM coupling.  

Numerical simulations were performed to help understand the MM Mamyshev oscillator. Drawing precise comparisons between experiment and simulation is difficult in STML lasers, but some features of a simulated cavity help explain the MM pulse evolution in this oscillator and align qualitatively with our experimental observations. To limit the time it takes to run simulations, we include up to 6 linearly polarized modes. The simulation \cite{gmmnlse} includes the effects of Kerr and Raman nonlinearities, self-steepening, and second- and third- order dispersion. We model the Yb gain by solving the rate equations in the steady state \cite{Turitsyn:11} \cite{RP}. To model coupling into and out of the MM arm, we assume a 1:1 imaging system between the end faces of the SM and MM fibers, and calculate overlap integrals of each MM fiber mode with the SM fiber mode. This includes spatial offset between fiber centers in order to excite or couple modes that are not radially-symmetric. Simulations of the same cavity with different input and output offsets show that the input offset primarily determines the modal energy content of the amplified MM pulse despite energy exchange between modes during propagation. This is consistent with our experimental observation that the SM to MM coupling controls the modal energy distribution. However, the general features of the evolution are similar for all input and output offsets up to 8.5 µm.

\begin{figure}[h!]
\centering
\includegraphics[width=\linewidth]{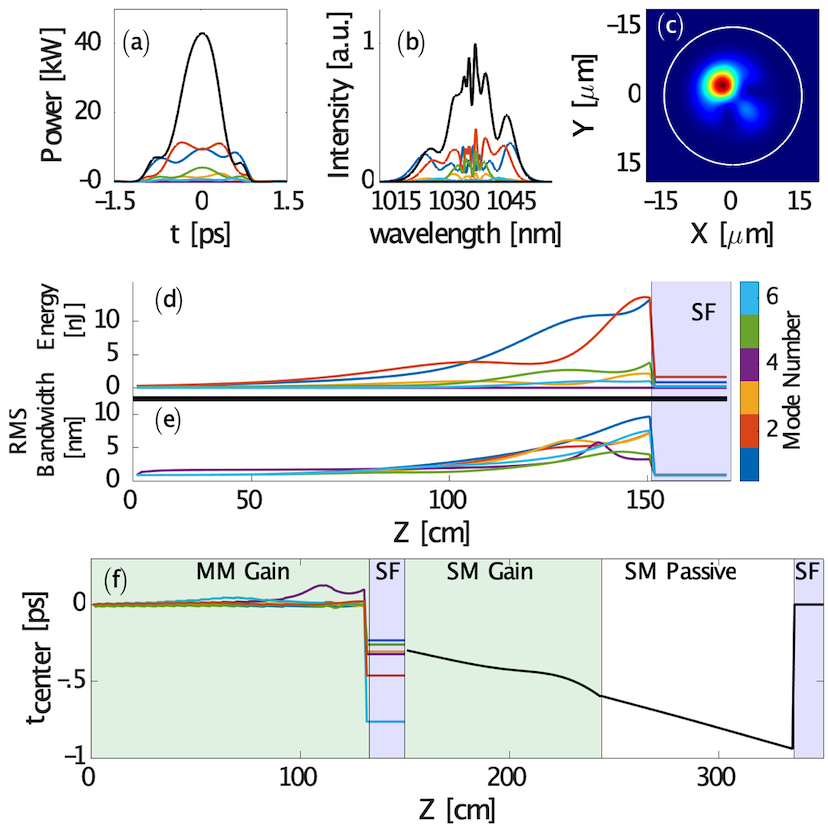}
\caption{Converged simulated MM pulse evolution. (a) Intensity of a spatially-integrated MM pulse (black) at the end of MM gain fiber, along with pulse in each mode with colors indicated by the colorbar in (d). (b) Power spectra of the spatially-integrated (black) pulse and individual modes (colored). (c) time-integrated spatial intensity as would be captured by a camera. (d) and (e) show energy and RMS bandwidth, respectively, per mode with evolution coordinate. The blue region captioned SF represents the spectral filter. (f) shows the time-center coordinate (first moment) of the MM and SM pulses for a complete cavity round-trip, relative to the simulation frame. Regions in green, blue, and white represent gain, spectral filtering, and passive propagation respectively. The time center coordinate is reset during each round trip. The colors used for the modes are the same as in (a) and (b).}
\label{fig:mmevol}
\end{figure}

Fig. \ref{fig:mmevol} shows various parameters of a typical simulated MM evolution for this cavity. Nonlinear spectral broadening occurs due to self-phase modulation (SPM) and cross-phase modulation (XPM) among multiple modes. XPM contributes more significantly to nonlinear phase than SPM \cite{Agrawal}\cite{Poletti:08}, so the spectral broadening  per mode is greater than it would be for SM propagation of pulses with the same energy. As shown in Fig. \ref{fig:mmevol}(e) and (d), pulses in several modes spectrally broaden sufficiently to reach the offset spectral filter, despite having little of the overall MM pulse energy. 

Previous work on STML lasers \cite{Wright94}\cite{Wright2020} emphasized the importance of spatial and spectral filtering to reset temporal walkoff between modes owing to intermodal group velocity differences. Our simulations indicate that the spectral filter following the MM arm does not decrease the temporal walkoff between pulses in different modes. Instead, temporal walk-off that arises from MM propagation and spectral filtering is corrected by the spatial filtering that occurs on coupling into the SM fiber (Fig. \ref{fig:mmevol}(f)).


In general, MM oscillators and amplifiers are not yet understood well enough to be fully controlled or employed as practical devices. The results presented here are a first step toward filling this knowledge gap for the case of oscillators or regenerative amplifiers that employ the Mamyshev mechanism. The large space of spatiotemporally-mode-locked states should be a fertile ground for scientific exploration. Future goals will include demonstration of control over the modal content of such lasers, scaling of the pulse energy to much-higher values, and generation of high-energy pulses with high-quality beams. 

In conclusion, we have constructed and characterized an STML laser with the Mamyshev mechanism as saturable absorber. The design features a SM arm in the cavity that supports the MM evolution by providing spatial filtering, compensating modal dispersion, and exerting some control over the modal content in the MM arm. The oscillator produces a wide variety of MM pulses with varying degrees of spatiotemporal coupling. Simulations and experiments indicate that nonlinear intermodal interactions, spatial filtering, and the Mamyshev mechanism together stabilize the evolution.

\begin{backmatter}
\bmsection{Funding} National Science Foundation (ECCS-1912742); Office of Naval Research (N00014-20-1-2789); Israeli Ministry of Defense (4441069183); Simons Foundation (733682).
\bmsection {Disclosures} The authors declare no conflicts of interest.
\bmsection{Data Availability} Data underlying the results presented in this paper are not publicly available at this time but may be obtained from the authors upon reasonable request.
\end{backmatter}

\bibliography{bibliography}

\bibliographyfullrefs{bibliography}

\end{document}